\documentstyle[12pt]{article}

\newcommand{\eq}{\begin{equation}}
\newcommand{\en}{\end{equation}}
\newcommand{\bea}{\begin{eqnarray}}
\newcommand{\eea}{\end{eqnarray}}

\newcommand{\wave}[2]{\mbox{\raisebox{-.6ex}{$\stackrel{\displaystyle{\sim}}
                     {\scriptstyle{{#1} \rightarrow {#2}}}$}}}

\newcommand{\D}{\Delta}

\newcommand{\NP}[1]{Nucl.\ Phys.\ {\bf #1}}
\newcommand{\PL}[1]{Phys.\ Lett.\ {\bf #1}}

\newcommand{\CMP}[1]{Comm.\ Math.\ Phys.\ {\bf #1}}

\newcommand{\PRL}[1]{Phys.\ Rev.\ Lett.\ {\bf #1}}
\newcommand{\MPL}[1]{Mod.\ Phys.\ Lett.\ {\bf #1}}

\begin{document}

\renewcommand{\thefootnote}{\fnsymbol{footnote}}

\newpage
\setcounter{page}{0}

\vskip 1cm
\begin{center}
{\bf RG FLOWS OF NON-DIAGONAL MINIMAL MODELS PERTURBED BY $\phi_{1,3}$}\\
\vskip 1.8cm
{\large F.\ Ravanini}\\
\vskip .7cm
{\em Service de Physique Th{\'e}orique, C.E.A. - Saclay \footnote{
     Laboratoire de la Direction des Sciences de la Mati{\`e}re du
     Commissariat \`a l'Energie Atomique} \\
     Orme des Merisiers, F-91191 Gif-sur-Yvette, France\\
     and\\
     I.N.F.N. - Sez. di Bologna, Italy}
\end{center}
\vskip 1cm

\renewcommand{\thefootnote}{\arabic{footnote}}
\setcounter{footnote}{0}

\begin{abstract}
Studying perturbatively, for large $m$, the torus partition function of
both $(A,A)$ and $(A,D)$ series of minimal models
in the Cappelli, Itzykson, Zuber classification,
deformed by the least relevant operator $\phi_{(1,3)}$,
we disentangle the structure of $\phi_{1,3}$ flows.
The results are conjectured on reasonable ground to
be valid for all $m$. They show that $(A,A)$ models always flow to $(A,A)$
and $(A,D)$ ones to $(A,D)$. No hopping between the two series is possible.
Also, we give arguments that there exist 3 isolated flows $(E,A)\rightarrow
(A,E)$ that, together with the two series, should exhaust all the possible
$\phi_{1,3}$ flows.
Conservation (and symmetry breaking) of non-local currents along the flows
is discussed and put in relation to the $A,D,E$ classification.
\end{abstract}
\vskip .3cm
\begin{flushright}
Saclay preprint SPhT/91-147\\
October 1991
\end{flushright}
\vskip .3cm
Submitted for publication to Phys. Lett. B
\newpage

{\bf 1 -} A well established result
in Conformal Field Theory (CFT) is that, perturbing
the $m$-th {\em diagonal} minimal model (where the level
$m$ parametrizes the central
charge as $c=c(m)=1-\frac{6}{m(m+1)}$, $m=3,4,5...$)
by its least relevant operator, namely the scalar primary field $\phi_{1,3}$,
for positive values of the
perturbing coupling constant one moves along a Renormalization Group (RG)
flow that ends at the $(m-1)$-th diagonal minimal model~\cite{zam,lc},
while for negative values of the coupling a massive theory is defined,
whose scattering matrix is known~\cite{blc}.

However, all the investigations so far on this subject have not considered
the {\em non-diagonal} minimal models, i.e. the models labelled by $(A,D)$
in the Cappelli, Itzykson, Zuber ADE classification~\cite{ciz}.
$\phi_{1,3}$ perturbation of $(A,D)$ models are possible, as for all of them
a $\phi_{1,3}$ scalar operator surely exists in the local sector and is
unique \footnote{Except for the $(A_4,D_4)$ theory where there are two
$\phi_{1,3}^{\pm}$ operators.}. It is not difficult to convince oneself (see
below) that also in the $(A,D)$ series $\phi_{1,3}$ perturbations with positive
coupling induce a flow from level $m$ to level $m-1$ models. In general at
level $m-1$ there are two models: $(A,A)$ and $(A,D)$ \footnote{At levels
11, 12, 17, 18, 29, 30 there is a third {\em exceptional} model $(A,E)$.}.
The question arises: does $(A,D)$ at level $m$ flow to $(A,D)$ or to $(A,A)$
at level $m-1$?
and what is the role of $(A,E)$ models? is there any classification scheme for
these flows? In the following we try to give an answer to these questions.

We consider a CFT (that we call ``ultraviolet'' (UV)), belonging to the set
of minimal models, with action $S_{UV}$ and central charge $c_{UV}=c(m)$,
and perturb it by its least relevant
scalar operator $\phi_{1,3}(z,\bar{z})$.
The (euclidean) action of the perturbed model is then
\eq
S = S_{UV} + g\int dzd\bar{z} \phi_{1,3}(z,\bar{z})
\label{action}
\en
In lagrangian formalism the partition function is expressed in terms of
a functional integral
\eq
Z(g)=\int {\cal D}\phi e^{-S[\phi]}
\label{part}
\en
where $\phi$ is a set of ``fundamental'' fields, whose nature is not important
in the following. This quantity, divergent on the sphere, is well defined
on a toroidal geometry $T$. Substituting (\ref{action}) in (\ref{part}) and
developing to first order in the perturbing parameter $g$
\eq
Z(g)=Z_{UV} - g\int_T d^2w\langle \phi_{1,3}(w,\bar{w})\rangle_T + O(g^2)
\en
where $w,\bar{w}$ are coordinates on the torus $T$. Translation invariance
ensures that one point functions on the torus are independent of the position.
$\int_T d^2w$ measures the area of the torus, thus leading to
\eq
Z(g)=Z_{UV} - g\tau_I \langle \phi_{1,3} \rangle_T + O(g^2)
\en
Of course both $Z_{UV}$ and $\langle \phi_{1,3}\rangle_T$ have a dependance
on the modular parameters $\tau,\bar{\tau}$. Here $\tau_I=\mbox{Im}\tau=
-\mbox{Im}\bar{\tau}$. A.Zamolodchikov~\cite{zam}, and Ludwig
and Cardy~\cite{lc}
have shown that the applicability of this perturbation expansion is suitable
for $m$ large, i.e. when the conformal dimension of the $\phi_{1,3}$ operator
$\D_{1,3}=1-\epsilon$, with $\epsilon=\frac{2}{m+1}$ tends to 1 and
$\phi_{1,3}$ becomes nearly marginal. $\epsilon$ can be then
seen as the parameter of an $\epsilon$-expansion around dimension 2.
Indeed, in this regime the Callan-Symanzyk $\beta$-function
\eq
\beta(g)=2\epsilon g-(\pi C_{(13)(13)}^{(13)}+O(\epsilon)) g^2+O(g^3)
\en
where the structure constant
$C_{(13)(13)}^{(13)}=4/\sqrt{3}+O(\epsilon)$ for both $(A,A)$ and $(A,D)$ (see
below),
shows a non-trivial zero, with infrared (IR) nature,
at the point $g^*=\frac{\sqrt{3}\epsilon}{2\pi}+O(\epsilon^2)$, which is still
in the perturbative region. At the IR point conformal invariance is restored.
Central charge $c$ can only decrease on the RG flow from UV to IR~\cite{cth},
and $c_{UV}-c_{IR}$ can be estimated, integrating the
$\beta$-function~\cite{zam,lc}, to be $12/m^3+O(\epsilon^4)$
in agreement with $c_{IR}=c(m-1)$.

{}From the arguments given in~\cite{zam,lc}
it is easy to infer that if the UV theory is in the $(A,A)$ series, the IR one
must also belong to that series. This fact has been recently confirmed by
a calculation of the perturbed partition function at one loop~\cite{ghs} that
we are going to reproduce and extend for the non-diagonal case in what follows.
If however the UV model is in the $(A,D)$ series, from the previous discussion
and the results of~\cite{zam,lc} it is not clear if it would flow to the IR
$(A,D)$ or to the $(A,A)$ model at $m-1$. To disentangle between these two
possibilities one could study wandering of non scalar operators, which can be
cumbersome, or turn to the analysis of the torus partition function along the
lines of a recent paper of Ghoshal and Sen~\cite{ghs}. Here we shall adopt this
second strategy. As $Z_{IR}=Z(g^*)$, the central object in our calculation
is the difference $\delta Z=Z_{UV}-Z_{IR}$, that is given, at one
loop, by the expression
\eq
\delta Z=\frac{\sqrt{3}\epsilon\tau_I}{2\pi}\langle\phi_{1,3}\rangle_T
+O(\epsilon)
\en
where $\langle\phi_{1,3}\rangle_T$ can contain terms proportional to $1/
\epsilon$.
\vskip .5cm
{\bf 2 -} First of all, we have to compute the difference between the
candidate UV and IR partition functions in order to compare them later with
the one loop calculation. The most economic way to do this, as cleverly
suggested in~\cite{ghs}, is to use the formulas in~\cite{difsz} that map the
minimal model partition functions on the torus into combinations of the
gaussian ones $Z(R)$, $R$ being the compactification radius of the gaussian
free boson. As we only need to analyze the large $m$ limit, we can make use
of the asymptotics $Z(R) \wave{R}{\infty} \sqrt{2}R Z_0$ where
\eq
Z_0 = \frac{1}{\sqrt{\tau_I}|\eta(\tau)|^2}
\en
The calculation for the $(A,A)$
series has already been performed in~\cite{ghs}. The results for the cases
we need to analyze can be summarized as follows:
\bea
\mbox{for}~(A,A)\rightarrow(A,A) &\Rightarrow& \delta Z=\frac{1}{2}Z_0\\
\mbox{for}~(A,D)\rightarrow(A,D) &\Rightarrow& \delta Z=\frac{1}{4}Z_0\\
\mbox{for}~(A,D)\rightarrow(A,A) &\Rightarrow& \delta Z=\frac{m}{2}Z_0
\eea
Hence these three situations can be disentangled by the different coefficients
of $Z_0$ in $\delta Z$.
It is interesting to notice that in the last case $\delta Z$ is divergent as
$m$ grows, thus signaling that perturbation analysis may not be applicable
to this
situation. This is not important, as the calculation that follows will show
that this possibility is ruled out in favour of the second one.
\vskip .5cm
{\bf 3 -} The main object to compute in order to estimate $\delta Z$ is then
the one point function on the torus of the operator $\phi_{1,3}$ at the UV
point. Expressions for this function are known by various arguments~\cite{1pt},
but they are in general quite cumbersome to use. Luckily when $\D_{1,3}
\rightarrow 1$, Ghoshal and Sen~\cite{ghs} have shown by a simple argument
(that can be extended from $(A,A)$ to $(A,D)$ with no substantial
modification) that the final expression of such a one point function simplifies
to
\eq
\langle\phi_{1,3}\rangle_T=4\pi^2 \sum_{F\in{\cal A}} C_{(13),F}^F \chi_h
\chi_{\bar h}^*
\label{1pt}
\en
where $\chi_h(\tau)$ are the characters
of the Virasoro irreducible module of highest weight $h$ and
the sum is performed over the set ${\cal A}$ of all primary fields $F=(h,
\bar{h})$ in the modular invariant sector of the model.
This set ${\cal A}$ (i.e. the fusion algebra) is known to have
different characterizations for $(A,A)$
and $(A,D)$ series. Also, the structure constants $C_{(13),F}^F$ can take
different values in the $(A,A)$ and $(A,D)$ cases respectively. For $(A,A)$
models, primary fields are all scalar and labelled by a couple of integers
$(r,s)$ such that $1\leq r \leq m-1$ and $1\leq s \leq m$, together with
a doubling: $(r,s)=(m-r,m+1-s)$. Hence
\eq
\mbox{for}~(A,A)~\Rightarrow~\sum_{F\in{\cal A}} = \frac{1}{2}\sum_{r=1}^{m-1}
\sum_{s=1}^m
\en
For $(A,D)$ models~\cite{petk} the algebra ${\cal A}$ naturally acquires a
$Z_2$ grading and splits in two parts: ${\cal A}={\cal A}_0 \oplus {\cal A}_1$
such that ${\cal A}_0 \times {\cal A}_0 = {\cal A}_1 \times {\cal A}_1 =
{\cal A}_0$ and ${\cal A}_0\times {\cal A}_1={\cal A}_1$. The ${\cal A}_0$
subalgebra is composed of scalar primary fields, while the ${\cal A}_1$ set
contains scalars as well as bosonic fields of higher spin.
\eq
\begin{array}{lll}
\mbox{for $m$ odd,}~(A_{m-1},D_{\frac{m+3}{2}}) &\Rightarrow&
\left\{
\begin{array}{ll}
{\cal A}_0=&\{\phi_{r,s}|\D_{rs}=\bar{\D}_{rs}~\mbox{and $s$ odd}\} \\
{\cal A}_1=&\{\phi_F|\D_F=\D_{rs},\bar{\D}_F=\D_{r,m+1-s}\\
           &\mbox{and $s=(m+1)/2 \bmod 2$}\}
\end{array}\right. \\
\mbox{for $m$ even,}~(D_{\frac{m}{2}+1},A_m) &\Rightarrow&
\left\{
\begin{array}{ll}
{\cal A}_0=&\{\phi_{r,s}|\D_{rs}=\bar{\D}_{rs}~\mbox{and $r$ odd}\} \\
{\cal A}_1=&\{\phi_F|\D_F=\D_{rs},\bar{\D}_F=\D_{m-r,s}\\
           &\mbox{and $r=m/2 \bmod 2$}\}
\end{array}\right.
\end{array}
\en
This allows to specify the sums in eq.(\ref{1pt}) in the various cases
\begin{itemize}
\item for $(A_{m-1},A_m)$
\eq
\langle\phi_{1,3}\rangle_T=2\pi^2\sum_{r=1}^{m-1}\sum_{s=1}^m
C_{(13)(rs)}^{(rs)} |\chi_{rs}|^2 + O(\epsilon)
\label{aa}
\en
\item for $(A_{m-1},D_{\frac{m+3}{2}})$, $m$ odd
\eq
\begin{array}{ll}
\langle\phi_{1,3}\rangle_T & =2\pi^2\displaystyle{\sum_{r=1}^{m-1}\sum_{
\stackrel{s=1}{s~{\rm odd}}}^m
C_{(13)(rs)}^{(rs)} |\chi_{rs}|^2} \\
& +2\pi^2\displaystyle{
\sum_{r=1}^{m-1}\sum_{\stackrel{s=1}{s=\frac{m+1}{2}\bmod 2}}
^m C_{(13),F}^F \chi_{rs}\chi_{r,m+1-s}^* + O(\epsilon)}
\end{array}
\label{ad}
\en
\item for $(D_{\frac{m}{2}+1},A_m)$, $m$ even
\eq
\begin{array}{ll}
\langle\phi_{1,3}\rangle_T & =2\pi^2\displaystyle{
\sum_{\stackrel{r=1}{r~{\rm odd}}}^{m-1}
\sum_{s=1}^m C_{(13)(rs)}^{(rs)} |\chi_{rs}|^2} \\
& +2\pi^2\displaystyle{\sum_{\stackrel{r=1}{r=\frac{m}{2}\bmod 2}}^{m-1}
\sum_{s=1}^m
C_{(13),F}^F \chi_{rs}\chi_{m-r,s}^* + O(\epsilon)}
\end{array}
\label{da}
\en
\end{itemize}

The structure constants $C_{(13)(rs)}^{(rs)}$ for the $(A,A)$ series
can be deduced from the general expressions of Dotsenko and Fateev~\cite{df}
\eq
C_{(13)(rs)}^{(rs)}=-\frac{1}{2\sqrt{3}} \frac{\Gamma\left(\frac{m+\lambda}
{m+1}\right)\Gamma\left(\frac{m-\lambda}{m+1}\right)}
{\Gamma\left(\frac{1+\lambda}{m+1}\right)\Gamma\left(\frac{1-\lambda}{m+1}
\right)}
\label{struct}
\en
where $\lambda=(m+1)r-ms$.
For the $(A,D)$ series the structure constants have been given by
Petkova~\cite{petk} and, as far as only field in the ${\cal A}_0$ subalgebra
are concerned, they coincide with those of the $(A,A)$ series. We shall see
that we need indeed only this subset of $(A,D)$ structure constants; for the
interested reader, a nice expression for the remaining ones can be found
in~\cite{petk}.
\vskip .5cm
{\bf 4 -} The other delicate point is the analysis of asymptotics of
characters for large $m$. Also this problem has been carried out in~\cite{ghs}
for the $(A,A)$ case. The general expression for a Virasoro character is
\eq
\chi_{rs}(\tau)=\frac{K_{r,s}(\tau)-K_{r,-s}(\tau)}{\eta(\tau)}
\en
where $\eta(\tau)$ is the Dedekind function and
\eq
K_{r,s}(\tau)=\sum_{k=-\infty}^{+\infty}
e^{2\pi i \tau\frac{[2m(m+1)k+(m+1)r-ms]^2}{4m(m+1)}}
\en
For $m$ large, terms are exponentially suppressed in $K_{r,s}$ unless $k=0$
and $r-s\sim O(1)$. The
other function $K_{r,-s}$ is made of exponentially
suppressed terms, unless $k=0$ and both $r,s\sim O(1)$, or $k=-1$ and both
$r,s\sim O(m)$. The set of characters that take contributions
from $K_{r,-s}$ is then of measure zero (in the large $m$ limit) compared
to the whole rectangle $1\leq r \leq m-1$, $1\leq s \leq m$. Hence
contributions to the sum over this rectangle from the second piece can be
neglected in the large $m$ limit. This analysis is sufficient to carry out
the calculation of $\langle\phi_{1,3}\rangle_T$ for the $(A,A)$
case~\cite{ghs}, namely taking as an expression of the character for large
$m$ the following
\eq
\chi_{rs}=\frac{1}{\eta}e^{\pi i \tau\frac{\lambda^2}{2m(m+1)}}
\label{char}
\en

What is surprising is that in the $(A,D)$ case, the contributions from the
mixed terms are always suppressed. Indeed, in the off-diagonal terms the
characters $\chi_{r,m+1-s}$ for $m$ odd or $\chi_{m-r,s}$ for $m$ even appear.
Take $m$ odd to fix the ideas. Terms are exponentially suppressed in
$K_{r,m+1-s}$ unless $k=0$ and $r-s+m\sim O(1)$, and in $K_{r,-(m+1)+s}$
unless $k=0$ and $r\sim O(1)$, $s\sim O(m)$, or $k=-1$ and $r\sim O(m)$,
$s\sim O(1)$. Thus $K_{r,-(m+1)+s}$ gives contibutions from a set of measure
zero (for $m$ large) and can be neglected. The only term that may survive is
the product $K_{r,s}K_{r,m+1-s}^*$, which may have contributions only from the
intersection between the non suppressed $(r,s)$ region for $K_{r,s}$ and
that for $K_{r,m+1-s}$. This intersection, however, in the large $m$ limit
is of measure zero too, so that it can be discarded as negligible.
The contribution from the structure
constants in the non-diagonal terms can at maximum account for a polynomial
divergence in $m$, that is anyway suppressed by the exponential dump of
all the terms in the characters. The same argument
can be carried out for the $m$ even case. In conclusion, {\em for
large $m$ there is no contribution to $\langle\phi_{1,3}\rangle_T$ from the
terms non diagonal in the characters in the expressions} (\ref{ad},\ref{da}).
\vskip .5cm
{\bf 5 -} This fact allows to finally compute $\delta Z$ for both $(A,A)$
and $(A,D)$ series. All what we have to do is to collect expressions
(\ref{struct},\ref{char}) and use them into (\ref{aa},\ref{ad},\ref{da})
to evaluate the one point function of $\phi_{1,3}$ and hence $\delta Z$.
It is useful, instead of trying to sum on $r$ and $s$ separately, to express
the sums in terms of $\lambda$, i.e.
\eq
\langle\phi_{1,3}\rangle_T=-\frac{\pi^2}{\sqrt{3}|\eta|^2}
\sum_{\lambda}
\frac{\Gamma\left(\frac{m+\lambda}{m+1}\right)
\Gamma\left(\frac{m-\lambda}{m+1}\right)}
{\Gamma\left(\frac{1+\lambda}{m+1}\right)
\Gamma\left(\frac{1-\lambda}{m+1}\right)}
e^{-\pi\tau_I\frac{\lambda^2}{m(m+1)}} + O(\epsilon)
\en
The set of $\lambda$ over which the sum has to be performed is the
only difference between $(A,A)$ and $(A,D)$ cases:
\begin{itemize}
\item for $(A,A)$, $\lambda$ runs from $-m(m-1)$ to $m(m-1)$ with the
exclusion of all values that divide $m$ or $m+1$,
\item for $(A,D)$, $\lambda$ takes all {\em odd} values with the exclusion
of those that divide $m$ or $m+1$. Indeed for $m$ odd, $s$ is always odd in
(\ref{ad}), and $(m+1)r-ms = \mbox{even}-\mbox{odd}=\mbox{odd}$. Similarly
for $m$ even $(m+1)r-ms=\mbox{odd}-\mbox{even}=\mbox{odd}$.
\end{itemize}
Next, for $(A,A)$ models, we can transform the sum into an integral by
considering the variable
$x=\lambda/(m+1)$, which, for large $m$,
is dense in its integration interval
$(-\infty,+\infty)$. The values of $\lambda$ dividing $m$ or
$m+1$ become a set of measure zero in the variable $x$ and this constraint can
be forgotten. Finally, remembering that $\Gamma(1+x)=x\Gamma(x)$, for $(A,A)$
we have the integral
\eq
\delta Z=\frac{\pi\epsilon}{4|\eta|^2}\frac{1}{\epsilon}\int_{-\infty}^
{+\infty}dx x^2 e^{-\pi\tau_I x^2}+O(\epsilon)=\frac{1}{2}Z_0+O(\epsilon)
\en
For $(A,D)$ as $\lambda$ takes odd values only, it is convenient to parametrize
$\lambda=2\rho-1$ and then define $x=\rho/(m+1)$, therefore
\eq
\delta Z=\frac{\pi\epsilon}{4|\eta|^2}\frac{1}{\epsilon}\int_{-\infty}^
{+\infty}dx 4x^2e^{-4\pi\tau_I x^2}+O(\epsilon)=\frac{1}{4}Z_0+O(\epsilon)
\en
Hence, comparing with the results of sect.2, we can state that
{\em for $m$ large, $\phi_{1,3}$
perturbations can only let $(A,A)$ flow to $(A,A)$ and $(A,D)$ flow to
$(A,D)$.} The picture for large $m$ then shows two series
between which there is no possible bridge created by a $\phi_{1,3}$ flow.
In our opinion, this result can be reasonably conjectured to be true {\em
for all $m$}, as we can not see any reason why a particular value of $m$
might be selected for which this picture breaks.
When the lowest $(A,D)$ model is reached, namely at $m=5$
(3-state Potts model $(A_4,D_4)$), we could still imagine a flow along the
``$(A,D)$'' series to the $(D_3,A_4)$ model. The coincidence $D_3=A_3$ tells
us that this model has to be identified with that of the $(A,A)$ series,
namely the tricritical Ising model $(A_3,A_4)$ at $m=4$.
We then conclude that at $m=5$ the
series of $(A,D)$ $\phi_{1,3}$ flows finally converges into the $(A,A)$ one.
This flow from 3-Potts to tricritical Ising, that should not be confused with
the flow $(A_4,A_5)\rightarrow(A_3,A_4)$, has been recently described by
Fateev and Al.Zamolodchikov~\cite{falz}.
\vskip .5cm
{\bf 6 -} Assuming that the picture just described is valid in general, an
intriguing observation comes to the eye. The $\phi_{1,3}$ flows among minimal
models are such that one of the two Lie algebras labelling the UV model in
the ADE classfication of~\cite{ciz} is still present in the IR one.
The other algebra has been substituted by a lower rank one. More precisely,
if we order the two Lie algebras labelling a model such that the first one has
lower dual Coxeter number, the rule is that the first Lie
algebra of the UV model always
coincides with the second algebra of the IR one. The IR model then remembers
part of the structure of the partition function of the UV one, and this is
a clear signal that there is some ``structure'' conserved along the flow.
This structure can be easily identified with the algebra of non-local
conserved currents described in~\cite{blc}. Each minimal model has two (left)
algebras of conserved currents, generated through operator product expansion
by the non-local operators $\psi(z)$ and $J(z)$
of conformal dimension $(\D_{13},0)$ and
$(\D_{31},0)$ respectively (corresponding right algebras are generated by
$(0,\D_{13})$ and $(0,\D_{31})$). The two indices labelling primary fields
in minimal models are related to the two different $\widehat{SU(2)}$
structures appearing in the Goddard, Kent, Olive (GKO)~\cite{gko} construction.
The first (resp. second) of the two ADE algebras in the classification scheme
is in relation with the first  (resp. second) $\widehat{SU(2)}$ structure,
hence with the first (resp. second) index of primary fields $r$ (resp. $s$).
The perturbing field $\phi_{1,3}$ behaves like an identity w.r.t. the first
$\widehat{SU(2)}$ structure and breaks explicitly only the second one. This
explains why the first ADE algebra is still present at the end of the flow.
The non-local algebra associated to this first structure is the one
generated by the current $J(z)$ and indeed this is shown in~\cite{blc} to be
conserved off criticality too, at least by perturbative arguments. A well
known result for $(A,A)$ that can trivially be extended to $(A,D)$ is that
the fields with left conformal dimension $\D_{31}$ at UV evolve to fields
with left conformal dimension $\D_{13}$ at IR along a $\phi_{1,3}$ flow.
This means that the $J(z)$ current of the UV point evolves towards the
$\psi(z)$ field of the IR point. The algebra generated by $\psi(z)$ is
associated with the second index $s$ of primary fields and with the second
algebra in the ADE classification. This clarifies why the first algebra at
UV becomes the second at IR.

To be more precise, instead of speaking of conservation of $J(z)$ along the
flow, it would be better to think in terms of spontaneous symmetry breaking.
Indeed, as widely documented~\cite{kms,zamprep} in the literature for the
simplest cases of tricritical Ising flowing to Ising and tricritical 3-state
Potts flowing to 3-state Potts, the UV conserved current $J(z)$ ensures a
symmetry of the theory which is no more present in the IR limit. For example
the $J(z)$ current for tricritical Ising is the fermionic partner of
stress-energy tensor, guaranteeing supersymmetry of the UV point. The IR point
(Ising model) is not supersymmetric, and the $\psi(z)$ field (the Onsager
fermion) has been
interpreted as the goldstino of the spontaneously broken supersymmetry. We
think that this picture can be extended to the general case: along the
$\phi_{1,3}$ flows there is spontaneous symmetry breaking of the symmetry
encoded in the non-local algebra generated by the $J(z)$ current. The
corresponding goldstino evolves to the $\psi(z)$ field as the IR limit is
approached. Notice that the sum of the left conformal dimension of the
spontaneously broken current at UV and of the goldstino field at IR always
equals 2, and this intriguing observation seems to be a general feature of two
dimensional spontaneous symmetry breaking to be better understood.

Tricritical Ising can also be seen as the lowest of a series
of minimal models for the superconformal algebra generated by stress-energy
tensor and the $J(z)$ current. It is known that ``fractional''
superconformal algebras can be generated by the stress-energy tensor plus
$J(z)$ currents in general~\cite{rav,blc,lct}
and show series of minimal models that can be identified
with $SU(2)\otimes SU(2)/SU(2)$ GKO cosets. Flows
between models in these series  have been studied, at least for $(A,A)$
series~\cite{sotkov}.
The lowest model of each series is a minimal model (in the usual Virasoro
sense), and if we insist to perturb it by $\phi_{1,3}$ we go to an IR limit
that does not belong to the series of $J(z)$ invariant models any more, thus
forcing spontaneous symmetry breaking of the fractional supersymmetry.

Finally, the (spontaneously broken) conservation of the current $J(z)$ allows
us to complete the picture of $\phi_{1,3}$ flows between minimal models, by
considering the $(A,E)$ models too. Let us concentrate on $E_6$ (the same
arguments can be repeated for $E_7$ and $E_8$). There are two $(A,E_6)$
models, namely $(E_6,A_{12})$ at $m=12$ and $(A_{10},E_6)$ at $m=11$. Let us
first consider the $(E_6,A_{12})$ model as an UV point. The $\phi_{1,3}$
operator exists in the model and can be shown to generate a perturbation
whose $\beta$-function is of the same form as for $(A,A)$ series. Hence we
reasonably expect that it flows to an IR model at $m=11$. There are 3 models
at $m=11$, but if we further assume that the first ADE algebra of UV must
coincide with the second of IR, we uniquely select the $(A_{10},E_6)$ model.
Thus this argument strongly suggests that there is a flow $(E_6,A_{12})
\rightarrow (A_{10},E_6)$. Next one can ask if the $(A_{10},E_6)$ model can
be considered as an UV point of a $\phi_{1,3}$ flow. The answer is {\em no},
for the simple reason that there is no local scalar operator $\phi_{1,3}$ in
this model, as an inspection of the modular invariant partition function shows.
Hence we cannot even speak of $\phi_{1,3}$ perturbation of this model.
The other question is if it exists a model at $m>12$ that can flow
to $(E_6,A_{12})$. This model should contain $A_{12}$ as its first Lie
algebra, hence the only candidates are $(A_{12},A_{13})$ or $(A_{12},D_8)$ at
$m=13$. Both are ruled out simply by the fact that the
$\phi_{1,3}$ flow defines a theory with no ambiguity, i.e. it must have
a unique IR limit, which is $(A_{11},A_{12})$ and $(D_6,A_{12})$
respectively, living no room for $(E_6,A_{12})$ as an IR limit of a
$\phi_{1,3}$ flow. Hence we conclude that, besides the two series $(A,A)
\rightarrow (A,A)$ and $(A,D)\rightarrow (A,D)$ there exist only 3 other
isolated flows $(E,A)\rightarrow (A,E)$ and this should reasonably exhaust
all the possible $\phi_{1,3}$ flows between minimal models, as
summarized in the following table:
\[
\begin{array}{lll}
\mbox{for all $m$:} & (A_{m-1},A_m)\rightarrow (A_{m-2},A_{m-1})
                    & \mbox{conserving $A_{m-1}$} \\
\mbox{for $m$ even:}& (D_{\frac{m}{2}+1},A_m)\rightarrow
                      (A_{m-2},D_{\frac{m}{2}+1})
                    & \mbox{conserving $D_{\frac{m}{2}+1}$} \\
\mbox{for $m$ odd:} & (A_{m-1},D_{\frac{m+3}{2}})\rightarrow
                      (D_{\frac{m+1}{2}},A_{m-1})
                    & \mbox{conserving $A_{m-1}$} \\
\mbox{for $m=12$:}  & (E_6,A_{12})\rightarrow (A_{10},E_6)
                    & \mbox{conserving $E_6$} \\
\mbox{for $m=18$:}  & (E_7,A_{18})\rightarrow (A_{16},E_7)
                    & \mbox{conserving $E_7$} \\
\mbox{for $m=30$:}  & (E_8,A_{30})\rightarrow (A_{28},E_8)
                    & \mbox{conserving $E_8$}
\end{array}
\]
Notice that the possible algebras of conserved currents (generated by $J(z)$)
follow an ADE classification too. Also notice that this list is in
agreementt with similar results on the perturbation of minimal models
{\em coupled to gravity} recently obtained using KdV techniques by Di Francesco
and Kutasov~\cite{dfk}.

Concluding, this paper only gives a perturbative argument
and makes use of some observation
concerning the conserved non-local currents to disentangle the structure of
$\phi_{1,3}$ flows between minimal models. We believe that a better
understanding of the complete picture of integrable perturbations of minimal
models, that should include for example knowledge of the scattering matrices
for $\phi_{1,3}$ perturbed $(A,D)$ and $(A,E)$
models in the negative coupling (massive)
direction, as well as correlation functions for non-critical models,
could be obtained by a deeper study of the non-local currents and
the quantum group symmetries underlying them.
\vskip .5cm
{\bf Acknowledgements -} I greatly acknowledge J.-B.Zuber for pointing me out
the paper~\cite{ghs}, which is the basis of the present work, for very
interesting discussions and for a careful reading of the manuscript. I am
grateful to D.Bernard for the continuous interest in this work and for many
useful and deep exchanges of ideas on non-local currents. The idea of studying
$\phi_{1,3}$ perturbation of $(A,D)$ models was born in conversations with
A.Cappelli. P.Di Francesco is also acknowledged for useful discussions.
I thank the Service de Physique Th{\'e}orique of C.E.A. - Saclay
for the kind hospitality and the Theory Group and the Director of Sez. di
Bologna of I.N.F.N. for the financial support allowing me to spend this year
in Saclay.

\end{document}